\documentclass[twocolumn,showpacs,preprintnumbers, superscriptaddress,eqsecnum,nofootinbib]{revtex4-1}
\usepackage{graphicx, fancybox}
\usepackage{amsmath,amssymb}
\usepackage[dvipdfm, 
pdfstartview=FitV, linkcolor=red, citecolor=blue, urlcolor=blue]{hyperref}
\usepackage{color}

\newcommand{\vect}[1]{\mathbf{#1}}

\newcommand{\re}{\mathrm{Re}}
\newcommand{\sgn}{\mathrm{sgn}}
\newcommand{\Slash}[1]{\ooalign{\hfil/\hfil\crcr$#1$}}
\newcommand{\vp}{\vect{p}}
\newcommand{\vk}{\vect{k}}

\newcommand{\vzero}{\vect{0}}
\newcommand{\vgamma}{{\boldsymbol \gamma}}
\newcommand{\valpha}{{\boldsymbol \alpha}}

\newcommand{\mb}{m_b}
\newcommand{\mf}{m_f}
\newcommand{\zetaf}{\zeta_f}
\newcommand{\zetab}{\zeta_b}

\newcommand{\deltam}{\delta m^2}

\newcommand{\comment}[1]{}

\newcommand{\R}{R}
\newcommand{\A}{A}
\newcommand{\cp}{g}
\newcommand{\nf}{n_F}
\newcommand{\nb}{n_B}
\newcommand{\Z}{Z}
\newcommand{\K}{L}

\newcommand{\const}{A}
\newcommand{\damping}{\varGamma}

\newcommand \beq{\begin{eqnarray}}
\newcommand \eeq{\end{eqnarray}}

\begin{document}

\title{Ultra-soft fermionic excitation at finite chemical potential}

\author{Jean-Paul Blaizot}
\email{jean-paul.blaizot@cea.fr} 
\affiliation{Institut de Physique Th\'eorique, CNRS/URA 2306, CEA/Saclay, F-91191 Gif-sur-Yvette Cedex, France}

\author{Daisuke Satow}
\email{daisuke.sato@riken.jp}
\affiliation{Theoretical Research Division, Nishina Center,
RIKEN, Wako 351-0198, Japan}
\affiliation{Department of Physics, Brookhaven National Laboratory, Upton, NY-11973, USA}

\begin{abstract} 
It has been suggested previously  that an ultra-soft fermionic excitation develops, albeit with a small spectral weight, in a system of massless fermions and scalar bosons with Yukawa interaction at high temperature $T$.
In this paper we study how this excitation is modified at finite chemical potential $\mu$. We relate the existence of the ultra-soft mode to symmetries, in particular charge conjugation, and a supersymmetry of the free system which is spontaneously broken by finite temperature and finite density effects, as argued earlier by Lebedev and Smilga.  
A non vanishing chemical potential breaks both symmetries explicitly, and maximally at zero temperature where the mode ceases to exist. A detailed calculation indicates that the ultra-soft excitation persists as long as  $T \geq 0.71\mu$.
\end{abstract}

\date{\today}

\pacs{
11.10.Wx, 
12.38.Mh,	
52.27.Ny	
}
\maketitle

\section{Introduction}
\label{sec:intro}

\begin{figure}[t] 
\begin{center}
\includegraphics[width=0.5\textwidth]{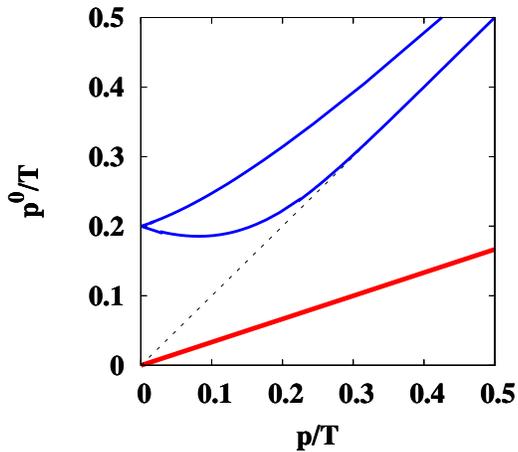} 
\caption{The dispersion relation of soft and ultra-soft excitations with positive energies (the spectrum of negative energies is the mirror image of this one) at $T\neq 0$, $\mu=0$, $\cp=0.8$. 
The upper two branches (solid line, blue) correspond to the plasmino, with the upper branch going into the normal fermion state at high momentum, and the lower branch disappearing from the spectrum when $p\gg gT$.
The lower branch (thick solid line, red) corresponds to the ultra-soft fermionic excitation.
Let $\valpha\equiv\gamma_0\vgamma$. 
The ultra-soft branch, as well as the lower plasmino branch, corresponds to a negative eigenvalue of $\valpha\cdot\hat\vp$, while the upper plasmino branch corresponds to a positive eigenvalue. Thus, the lower two branches of the spectrum shown in this figure carry quantum numbers that are normally attributed to antiparticles.} 
\label{fig:dispersion-mode}
\end{center}
\end{figure}

The spectrum of quasi-particles or collective excitations is an important property of a many-body system. In systems with long range interactions, such as plasmas, such collective modes are well studied, in non-relativistic as well as relativistic cases, in electrodynamics (QED), and in quantum chromodynamics (QCD)~\cite{plasmon,HDL,HTL, blaizot-mu, plasmino, lebedev, Baym:1992eu, QED-ultrasoft, Satow:2013oya, S-Deq, 3peak, mitsutani}. 
Similar modes also exist in simpler systems, such as the one studied in this paper, composed of massless fermions and scalar bosons with Yukawa coupling. 
Various methods have been developed to study these systems, based on perturbation theory (the hard thermal/dense loop (HTL/HDL) approximation~\cite{plasmon, HTL,HDL}) or kinetic theory~\cite{blaizot-HTL}. 
One generically finds that collective phenomena develop on a momentum scale (referred to as soft scale) of order $\cp T$ or $\cp\mu$, where $\cp$ is the coupling constant, $T$ is the temperature, and $\mu$ the chemical potential controlling the fermion density.
A particularly noticeable feature is the splitting of the fermion spectrum in this momentum region, a phenomenon that has been referred to as the ``plasmino'' (see Fig.~\ref{fig:dispersion-mode} for an illustration). 

In this paper, we shall be concerned with the region of ultra-soft momenta, $p\lesssim g^2T, g^2\mu$, where a new type of excitation is expected to appear. 
This is confirmed by a variety of calculations performed at $T\ne 0$ and $\mu=0$~\cite{lebedev, QED-ultrasoft, Satow:2013oya, S-Deq, 3peak, mitsutani, susy-kinetic} (see again Fig.~\ref{fig:dispersion-mode} for an illustration).
The purpose of the present paper is to investigate how this picture is modified at finite density. The motivation for doing this is the following. A nice interpretation of the ultra-soft excitation has been provided by Lebedev and Smilga (in the case of gauge theories) in term of a supersymmetry of the free Lagrangian, the ultra-soft excitation being interpreted as the Nambu-Goldstone mode (called quasi-goldstino)~\cite{lebedev} associated with the spontaneous breaking of this supersymmetry by thermal effects~\cite{Das:1978rx}. A finite chemical potential reveals the existence of another important symmetry, that of charge conjugation, which turns out to play an essential role in understanding the emergence of the quasi-goldstino. A finite chemical potential breaks charge conjugation symmetry, as well as supersymmetry in an explicit way, and a careful study of how this affects the existence of the quasi-goldstino sheds a new light on the physics of this particular excitation. 

This paper is organized as follows:
In the next section, we review the essential features of the ultra-soft fermionic excitation at high temperature and vanishing density, identifying the various symmetries that play a role in its existence.
In the following section we present explicit calculations at finite temperature and finite density, using resummed one-loop perturbation theory.
The last section summarizes our conclusions.

\section{Soft and ultra-soft fermionic excitations at high temperature}
\label{sec:ultrasoft}

We start by recalling general features of collective fermionic excitations in a plasma made of massless fermions and massless bosons.

The dispersion relations of the fermionic excitations may be deduced from the fermion retarded propagator $S^\R(p^0,\vp)$,
the general structure of which follows from elementary symmetry considerations. 
Symmetry under parity ($S(p^0,-\vp)=\gamma_0 S(p^0,\vp)\gamma_0$), together with chiral symmetry\footnote{In QED/QCD, chiral symmetry exists as long as the bare fermion mass is negligible compared with $T$.
Note that in the Yukawa model, which we consider in this paper, chiral symmetry is not a symmetry of the interaction part of the Lagrangian (\ref{eq:lagrangian}). 
However, since the expectation value of the field $\phi$ vanishes in equilibrium, there is no leading order contribution to the fermion mass. 
Actually, within the HTL approximation and the resummed one-loop approximation, which we shall consider in this paper, one can verify explicitly that the self-energy $\varSigma$ has indeed the structure indicated in Eq.~(\ref{SigmaRabc}), i.e., no constant term (proportional to the unit matrix) appears.} ($\gamma_5 S(p^0,\vp)\gamma_5=-S(p^0,\vp)$), allow us to write:
\begin{align}
\begin{split} 
S^\R(p^0,\vp)&= -\frac{1}{\Slash{p}-\varSigma^R(p^0,\vp)}\\
&= -\frac{1}{2}\left(\frac{\gamma^0-\vgamma\cdot\hat{\vp}}{S_+(p^0,|\vp|)}
+\frac{\gamma^0+\vgamma\cdot\hat{\vp}}{S_-(p^0,|\vp|)}\right),
\end{split}
\end{align}
where the scalar functions,
\beq
\label{eq:propagator-decompositon}
S_{\pm}(p^0, |\vp|)\equiv 
p^0-a(p^0, |\vp|)\mp(|\vp|+b(p^0, |\vp|)),\nonumber\\
\eeq
do not depend on the direction of $\vp$ because of rotational invariance. 
Here $a$ and $b$ are related to the retarded fermion self-energy $\varSigma^R(p^0,\vp)$
by
\beq\label{SigmaRabc}
\varSigma^R(p^0,\vp)=a(p^0, |\vp|)\,\gamma_0+b(p^0, |\vp|)\,\vgamma\cdot\hat{\vp},
\eeq
where $b(p^0, |\vp|)$ vanishes when $|\vp|=0$. 
When the system is invariant under charge conjugation, which is the case when the fermion chemical potential vanishes, we have also~\cite{weldon-structure}
\beq
\label{eq:symmetry-S1}
S_-(p^0, |\vp|)&=-(S_+(-p^0{}^*, |\vp|))^*.
\eeq
We now specialize to vanishing 3-momentum $|\vp|=0$. Since, when $|\vp|=0$,  $S^R$ cannot depend on the direction $\hat\vp$ of the momentum, $
S_+(p^0, 0)=S_-(p^0, 0).
$
When combined with Eq.~(\ref{eq:symmetry-S1}), this yields
\begin{align}
S_+(p^0,0)&=-(S_+(-p^0{}^*,0))^*.
\end{align}
This conditions translates into (with $p^0$ real)
\begin{align} 
\label{eq:symmetry-sigma}
\begin{split}
\re \,a(-p^0 , 0)=-\re\, a(p^0 ,0),
\end{split}
\end{align}
that is $\re$ $a(p^0 , 0)$ is an odd function of $p^0$. Thus, unless ${\rm Re}\,a(p^0,0)$ is singular at $p^0=0$, it vanishes there, and this entails the existence of a pole of $S^R$ at $p^0=0$, as can be seen from Eq.~(\ref{eq:propagator-decompositon}). This corresponds to the ultra-soft fermionic excitation, as long as the imaginary part of $a(p^0 , 0)$ is small enough. 

Let us then investigate the behavior of $a(p^0,0)$ at small $p^0$ in some simple approximations, and within a specific model, the   Yukawa model, whose Lagrangian reads 
\beq
\label{eq:lagrangian}
{\cal L}= \frac{1}{2}(\partial^\mu\phi)^2+\overline{\psi}(i\Slash{\partial}-\cp\phi-\mu\gamma_0)\psi.
\eeq 
Here, $\phi$ and $\psi$ are the scalar and the fermion fields, respectively, and possible self interactions of the scalar field are ignored (they play no role in the present discussion). 
The chemical potential $\mu$ controls the net fermion number. 

\begin{figure*}[t]
\begin{center}
\includegraphics[width=0.9\textwidth]{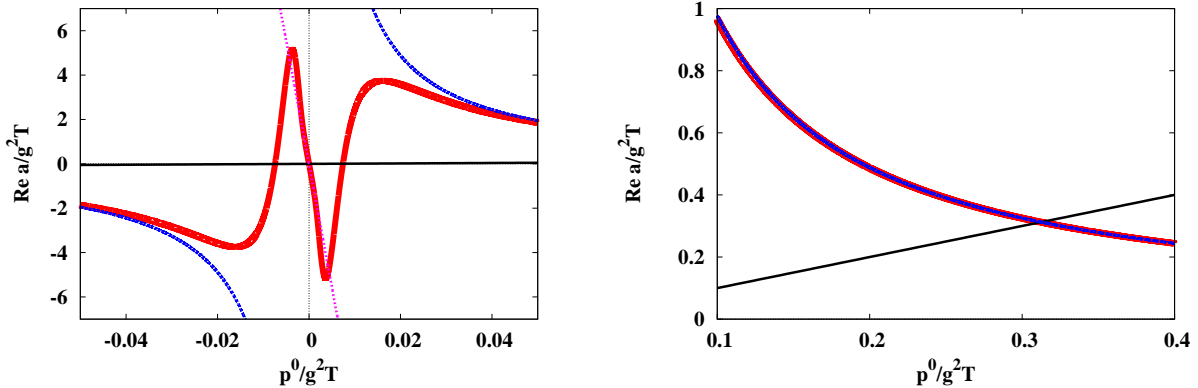} 
\caption{The real part of $a(p^0,0)$ as a function of $p^0$ at $T\neq 0$, $\mu=0$, and $\cp=0.8$ in the HTL approximation (dashed, blue), Eq.~(\ref{eq:HTL-a}), in the resummed one-loop approximation with $\zeta=0$ (thick solid, red), Eq.~(\ref{eq:sigma-mu}), and in the small $p^0$ expansion with $\damping=0$ (dotted, magenta), Eq.~(\ref{eq:expansion-ab}), for small $p^0$ (left panel) and large $p^0$ (right panel). 
Observe the change in the structure at small $p^0$ between the HTL result and the resummed perturbation result caused by the asymptotic masses. 
The pole position is determined by the intersection of the curves and the inverse free propagator (solid, black) $p^0$.
One pole is located at the plasmino frequency $p^0=\omega_0$.
The other pole corresponding to the ultra-soft excitation is located at $p^0=0$.  
There are also  two other solutions, but these are not physical since they are found in regions where the slope $\partial a(p^0,0)/\partial p^0$ is negative, corresponding to  residues larger than unity.
} 
\label{fig:plasmino-mode}
\end{center}
\end{figure*}

In the HTL approximation, we have 
\beq
\label{eq:HTL-a}
a(p^0, 0)=\frac{\omega_0^2}{p^0}
\eeq
where $\omega_0= gT/4$ is the plasmino frequency (the analog of the plasma frequency in ordinary plasmas). This is indeed an odd function of $p^0$, as expected, but it is not regular at $p^0=0$. 
In fact $a(p^0, 0)$ diverges as $\omega_0^2/p^0$ as $p^0\to 0$, as can be seen from Eq.~(\ref{eq:HTL-a}).
In this approximation, when $|\vp|=0$, there are poles at $p^0=\pm\omega_0$, but no pole exists around $p^0=0$. 
This behavior can be seen from the plot for the HTL result, Eq.~(\ref{eq:HTL-a}), in both panels of Fig.~\ref{fig:plasmino-mode}. 

The HTL approximation is valid only for soft external momenta $p\sim \cp T$, and it breaks down in the limit of ultra-soft momenta, $p\lesssim g^2 T$~\cite{lebedev, susy-kinetic, Satow:2013oya, blaizot-ultrasoft, transport-perturbation, transport-kinetic, transport-2PI, transport-3PI, QED-ultrasoft, ultrasoft-kinetic, blaizot-review}. 
The HTL approximation treats the hard particles as free, massless particles. 
An improved approximation, that will be presented explicitly in the next section, takes into account the thermal masses of the hard bosons and fermions (the so-called asymptotic masses \cite{scalar,gerhold-dispersion}), and this changes completely the structure of ${\rm Re}\, a(p^0,0)$ at small $p^0$. 
The function remains an odd function of $p^0$, but the divergence turns into a very rapid variation around $p^0=0$, the function vanishing linearly at $p^0=0$:  $a(p^0,0)\sim - p^0/Z$, where  $Z\sim g^2$ is small, hence the large negative slope. 
This behavior is illustrated  in the left panel of Fig.~\ref{fig:plasmino-mode} where we have plotted  the result of  resummed perturbation theory obtained from Eq.~(\ref{eq:sigma-mu}) below, together with the  small $p^0$ expansion given in Eq.~(\ref{eq:expansion-ab}).
Note that the intersections of the free propagator with the positive slope parts of ${\rm Re} \,a(p^0,0)$ do not correspond to physical excitations. This is because the residues of $S_\pm(p^0, 0)$ at the corresponding poles,  given by
\beq
\left(1-\frac{\partial a(p^0=x,0)}{\partial p^0}\right)^{-1} , 
\eeq
are larger than unity when the slope is positive. 
Note also that the modification of $a(p^0,0)$ with respect to the HTL approximation does not affect much the properties of the plasmino (the curves for the HTL result and the resummed perturbation in the right panel of Fig.~\ref{fig:plasmino-mode} are almost on top of each other in the vicinity of the crossing point with  the bare propagator): this is related to the fact that the  ultra-soft mode carries little ($\sim g^2$) spectral weight~\cite{QED-ultrasoft}.

\begin{figure}[t] 
\begin{center}
\includegraphics[width=0.3\textwidth]{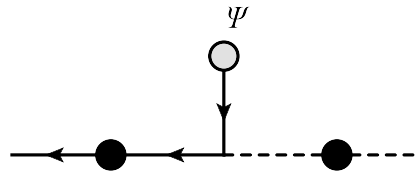}
\caption{Illustration of the process  that contributes to the ultra-soft fermion self-energy.
The solid (dashed) line with black blob is the resummed fermion (boson) propagator that contains the asymptotic thermal mass and the damping rate of the fermion (boson).
The ultra-soft fermion is treated as the average field ($\varPsi$), which is represented with the gray blob. 
The hard mode follows the horizontal line, and change from a boson to a fermion, or antifermion. 
A corresponding process where hard (anti)fermions turn into hard bosons also exists.  
When the masses of the boson and the fermion can be neglected, as in the HTL approximation, the process exhibits an apparent supersymmetry.
By contrast, this supersymmetry is broken when the mass difference between the fermion and the boson cannot be ignored, as is the case at low momenta.
}
\label{fig:kinetic}
\end{center}
\end{figure}

Another perspective on these behaviors can be gained by separating the degrees of freedom into soft (or ultra-soft) and hard (whose momentum is of order $T$) degrees of freedom. 
The collective excitations are associated with long wavelength, low frequency, oscillations of the soft degrees of freedom that can be represented by average fields, while the hard, particle,  degrees of freedom  account for the polarization of the medium by the soft fields \cite{blaizot-HTL, blaizot-review}. 
The processes that describe the response of the thermal medium to an applied soft field (here carrying fermionic quantum numbers) are displayed in  Fig.~\ref{fig:kinetic}. 
Such processes are encoded in a special propagator $K(x,y)=\langle {\rm T} \psi(x)\phi(y)  \rangle$ ({\rm T} is time-ordering operator) mixing  fermion and boson degrees of freedom, and which obeys the following generalized kinetic equation~\cite{blaizot-HTL, ultrasoft-kinetic}
\begin{align}
\label{eq:yukawa-kineticeq}
\begin{split}
&\left(2iv\cdot\partial_X\pm\frac{\delta m^2}{|\vk|}\right)\varLambda_\pm(\vk,X)\\
&\qquad=\cp\Slash{v}(\nb(|\vk|)+\nf(|\vk|))\varPsi(X),
\end{split} 
\end{align} 
where $v^\mu\equiv(1,\hat{\vk})$, $\hat{\vk}\equiv\vk/|\vk|$, $\nb$ ($\nf$) is the boson (fermion) distribution function, $\varPsi$ is the average  fermion field, and $\varLambda_\pm$ is the off-diagonal density matrix defined by $K(k,X)\equiv2\pi\delta(k^2)[\theta(k^0)\varLambda_+(\vk,X)+\theta(-k^0)\varLambda_-(-\vk,X)]$.
In Eq.~(\ref{eq:yukawa-kineticeq}), $\delta m^2\equiv \cp^2T^2/24$ is the difference between the square of the asymptotic thermal masses of the boson and the fermion, and  $K(k,X)$, with $X=(x+y)/2$, is the Wigner transform 
\beq
K(k,X)=\int {\rm d}^4s \,{\rm e}^{ik\cdot s} \,K\left(X+\frac{s}{2},X-\frac{s}{2}\right).
\eeq
For simplicity, we omit here the damping rate (which is of higher order in the coupling constant).
The equivalence between the kinetic description using Eq.~(\ref{eq:yukawa-kineticeq}) and the more standard diagrammatic calculation, which results in Eq.~(\ref{eq:sigma-mu}) below, follows from the following expression of the fermion self-energy~\cite{ultrasoft-kinetic, blaizot-HTL, blaizot-review}:
\begin{align}
\label{eq:selfenergy-expression}
\begin{split}
\varSigma^R(p)=\cp\int \frac{d^3\vk}{(2\pi)^3}\frac{1}{2|\vk|}\sum_{s=\pm} \frac{\delta \varLambda_{s}(\vk,p)}{\delta \varPsi(p)},
\end{split}
\end{align} 
where  $\varLambda_{\pm}(\vk,p)$ and $\varPsi(p)$ are the Fourier transforms of $\varLambda_{\pm}(\vk,X)$ and $\varPsi(X)$, respectively.

In the case of the soft mode (the plasmino), $p\sim \partial_X \sim gT$,  we can ignore the contribution of $\deltam$ in the l.h.s. of Eq.~(\ref{eq:yukawa-kineticeq}). 
This is equivalent to the HTL approximations, and the resulting self-energy is that given in Eq. (\ref{eq:HTL-a}) when $|\vp|=0$. 
Note that we have then $\varLambda_+=\varLambda_-$.
By contrast, when the momentum $p$ is ultra-soft ($p\sim\partial_X\ll \cp^2T$), the asymptotic thermal mass difference $\deltam$ dominates over the drift term in Eq.~(\ref{eq:yukawa-kineticeq}). 
In the limit where one neglects the drift term completely, $\varLambda_+=-\varLambda_-$, and from Eq.~(\ref{eq:selfenergy-expression}), one concludes that $\varSigma^R(0)$ vanishes. 
This discussion reveals the role of charge conjugation symmetry in the existence of the ultra-soft mode:  it is this symmetry that ensures  the cancellation between various processes such as that displayed in Fig.~\ref{fig:kinetic} and which eventually leads to the vanishing of $\varSigma^R(0)$. The discussion also points to the essential role of  $\delta m^2$, and this is best understood by referring to another symmetry, a supersymmetry.

The role of such a supersymmetry is already suggested by the nature of the physical processes that are responsible for the collective fermionic excitation, as  displayed in Fig.~\ref{fig:kinetic}: the main dynamics of hard particles involve turning fermions into bosons under the influence of a soft or ultra-soft field. 
If we consider the soft excitation, which enables us to use the free dispersion relations for the hard fermion and the hard  boson (HTL), there is degeneracy between these two particles, both being massless. 
At this level, the symmetry is reflected in the fact that hard bosons and fermions play identical roles, but does not entail any special additional consequence. 
Things are different in the region of ultra-soft momenta. 
As we have seen, the self-energy of the ultra-soft excitations is sensitive to the masses of the hard excitations, and bosons and fermions acquire different thermal masses: the degeneracy between hard  bosons and fermions is lifted, and this phenomenon is akin to a spontaneous symmetry breaking, with the massless ultra-soft excitation being  the associated Nambu-Goldstone mode. 
This interpretation was first proposed by Lebedev and Smilga~\cite{lebedev} in the case of QCD, but it extends trivially to the present case. 
Note that this scenario is explicitly realized in truly supersymmetric systems,  such as in the Wess-Zumino model, and the Nambu-Goldstone mode is there called goldstino~\cite{susy-kinetic}\footnote{In such models, supersymmetry also provides an interpretation for the velocity $v=1/3$ of the mode, with dispersion relation $p^0=\pm v |\vp|$. This velocity is the same in the Yukawa model or in QED/QCD~\cite{QED-ultrasoft, Satow:2013oya}, where it results from some angular integration, as that of the goldstino in the Wess-Zumino model at high temperature~\cite{susy-kinetic}.
In the latter case, one can argue that the goldstino being the superpartner of the phonon, sometimes called phonino, has its velocity given by $v=P/\epsilon$ ($\epsilon$: energy density, $P$: pressure), which is $1/3$ in a system of massless particles.}. 

In our case, supersymmetry is only approximate, and emerges only at high temperature where some interaction effects can be neglected. To be precise, consider the transformation of the fields defined by
\begin{align}
\delta \phi&= \overline{\eta}\psi+\overline{\psi}\eta,\\
\delta\psi&= -i \Slash{\partial}\phi\eta,\quad 
\delta\bar\psi= i\bar\eta\,\Slash{\partial}\phi
\end{align}
where $\eta$ is an infinitesimal Grassmann parameter. It is easy to see that this transformation leaves the free part of the Lagrangian (\ref{eq:lagrangian}) invariant when $\mu=0$, up to a total derivative (note  though that  the numbers of fermion and boson degrees of freedom  are different). 
The interaction term is not invariant and this will lead to explicit breaking terms that are however small at weak coupling. On the other hand, the thermal masses, that also result from interactions, lead, as we have seen, to non perturbative effects characteristic of spontaneous symmetry breaking. 
 
 We note also that the term proportional to the chemical potential breaks  explicitly  the supersymmetry (in addition to charge invariance). This effect can be large, and it is the purpose of the next section to analyze its effects on the ultra-soft excitation. 


\section{Ultra-soft excitation at finite chemical potential}
\label{sec:mu}

We have seen in the last section that the existence of the ultra-soft mode is related to the presence of two symmetries, an approximate supersymmetry that emerges at high temperature, and the charge conjugation symmetry. The former is broken spontaneously by temperature effects and this entails the existence of a soft mode by a mechanism analogous to the Nambu-Goldstone mechanism. The charge conjugation invariance on the other hand has been seen to be at the origin of important cancellations that are responsible for a particular structure of the self-energy that indeed allow the ultra-soft mode to develop, but it does not appear in itself a driving mechanism for the existence of the soft mode. It is then interesting to investigate situations where these symmetries are explicitly broken to some degree, and see whether the soft mode continues to exist then. This is the purpose of this section, where we use the specific Yukawa model introduced in the previous section, but now taking into account a finite chemical potential. The chemical potential breaks explicitly both the symmetries mentioned above, and indeed we shall see that at zero temperature, and finite chemical potential no soft mode appears. However,  at high temperature  the ultra-soft  mode continues to exist in the presence of a finite chemical potential, as long as the temperature stays  larger than the chemical potential, more precisely, as a detail analysis reveals, as long as $T\ge 0.71\mu$.  

We use the real time formalism~\cite{lebellac} throughout this section.
We analyze the fermion self-energy in the ultra-soft regime of momenta ($p\lesssim$ max($\cp^2T$, $\cp^2\mu$)). In this regime, special resummations are needed in order to avoid the pinch singularity that would appear~\cite{lebedev, susy-kinetic, Satow:2013oya, blaizot-ultrasoft, transport-perturbation, transport-kinetic, transport-2PI, transport-3PI, QED-ultrasoft, ultrasoft-kinetic, blaizot-review} in the HTL/HDL approximation. The appropriately resummed propagators at finite $T$ and $\mu$ have the form
\begin{align} \label{retprop0}
S^\R(k)&\approx -\frac{\Slash{k}}{k^2-\mf^2+2i\zetaf(k) k^0} , \\
D^\R(k) &\approx -\frac{1}{k^2-\mb^2+2i\zetab(k)k^0},
\end{align} 
with $S^\R(k)$  and $D^\R(k)$ respectively the fermion and boson retarded propagators. Note that the variable $k^0$ here measures the energy with respect to the chemical potential. These propagators will be needed in loop intergrals dominated by hard momenta $k\sim$ max$(T, \mu)$, and near the light-cone, $k^2\approx 0$ (see below). 
Accordingly $m_f$ and $m_b$ are the so-called `asymptotic masses' of the fermion and boson excitations~\cite{scalar, gerhold-dispersion}, whose expressions are
\beq
\mf^2= \frac{\cp^2}{8}\left(T^2+\frac{\mu^2}{\pi^2}\right),
 ~\mb^2= \frac{\cp^2}{6}\left(T^2+\frac{3\mu^2}{\pi^2}\right).
\eeq 
What will enter the calculation below is actually only the difference of these masses, 
\beq\label{asymmasses}
\deltam\equiv \mb^2-\mf^2=\frac{\cp^2 }{24}\left( T^2+\frac{9\mu^2}{\pi^2} \right).
\eeq
The quantities $\zetaf$ and $\zetab$ are the damping rates of the hard fermion and the boson. 
They are of order $\cp^4T$ $(\cp^4\mu)$ at large $T$ ($\mu$)  (up to a factor $\ln(1/\cp)$)~\cite{damping-mu, scalar}.
Since they are of higher order than the masses, they do not play an important role\footnote{This is to be contrasted with what happens in gauge theories, such as QCD/QED, where the damping is anomalously large~\cite{anomalous-damping}, $\zeta\sim g^2\ln (1/g)$, and the quasi-goldstino is over damped~\cite{lebedev, QED-ultrasoft, Satow:2013oya}.} in our discussion. In particular, we neglect them in making the plots for Figs.~\ref{fig:plasmino-mode}, \ref{fig:selfenergy-mu1}, and \ref{fig:selfenergy-mu2}.

\begin{figure}[t] 
\begin{center}
\includegraphics[width=0.3\textwidth]{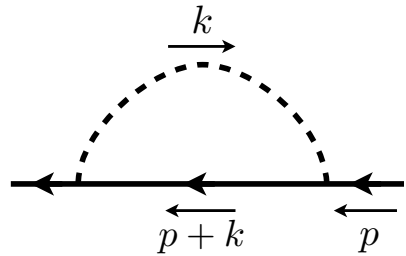}
\caption{Diagrammatic representation of the fermion self-energy in Eq.~(\ref{eq:sigma-mu}) 
at resummed one-loop order.
The solid and dashed lines are the resummed propagator of the fermion and the scalar boson, respectively.
} 
\label{fig:oneloop}
\end{center}
\end{figure}

The fermion retarded self-energy at the (resummed) one-loop order can be written as follows (see  Fig.~\ref{fig:oneloop}):
\begin{align}
\label{eq:sigma-mu0}
\begin{split}
\varSigma^\R(p)&= i\cp^2\int\frac{d^4k}{(2\pi)^4}[D^S(-k)S^\R(p+k) \\
&~~~+D^\R(-k)S^S(p+k)],
\end{split}
\end{align}
where $S^S(k)$ and $D^S(k)$ are proportional to the fermionic and bosonic spectral functions, respectively, 
\begin{align} 
\nonumber
 S^S(k) &\equiv \left(\frac{1}{2}-\nf(k^0)\right)(S^\R(k)-S^\A(k)) \\
 & \approx \left(\frac{1}{2}-\nf(k^0)\right)\Slash{k}\frac{4ik^0\zetaf(k) }{(k^2-\mf^2)^2+4(k^0)^2\zetaf^2(k)},\\
 \nonumber
 D^S(k) &\equiv \left(\frac{1}{2}+\nb(k^0)\right)(D^\R(k)-D^\A(k)) \\
& \approx \left(\frac{1}{2}+\nb(k^0)\right)\frac{4ik^0\zetab(k) }{(k^2-\mb^2)^2+4(k^0)^2\zetab^2(k)} ,
\end{align}
with $\nf(k^0)\equiv 1/\{\exp[(k^0-\mu)/T]+1\}$ and $\nb(k^0)\equiv 1/[\exp(k^0/T)-1]$. 
One could be worried that at the same order vertex corrections should also be included in addition to the resummed propagators, as in the case of QED/QCD~\cite{lebedev, QED-ultrasoft, Satow:2013oya}. 
However arguments similar to those used at $T\neq 0$, $\mu=0$~\cite{QED-ultrasoft, susy-kinetic}, show that this is not the case  for the Yukawa model considered here. 
A simple calculation, using the explicit expressions given above for the propagators and spectral functions, leads to 
\beq
\label{eq:sigma-mu}
\varSigma^\R(p)\approx \cp^2\int\frac{d^4k}{(2\pi)^4}\K(k) \frac{\Slash{k}}{\deltam+2k\cdot \tilde{p}}.
\eeq
where $\tilde{p}\equiv (p^0+i\zeta, \vp)$,  $\zeta\equiv\zetaf+\zetab$, and
\beq\label{defL}
 \K(k)\equiv 2\pi\sgn(k^0)\delta(k^2)(\nf(k^0)+\nb(k^0)).\eeq
In deriving (\ref{eq:sigma-mu}), we used the fact that $p$, $\mf$, $\mb$, $\zetaf $, $\zetab$ are much smaller than the typical loop momentum $k\sim {\rm max}(T, \mu)$, in order to drop certain terms. We also neglected vacuum contributions (that  do not depend on neither $T$ nor $\mu$). As anticipated, the momentum integral is dominated by hard momenta of on-shell quasi-particles (whose masses are neglected in the delta-function of Eq.~(\ref{defL})).
We note that the HTL/HDL approximation corresponds to neglecting the term $\deltam$ in Eq.~(\ref{eq:sigma-mu}), as already indicated. 
The corresponding result at $|\vp|=0$ is still given by Eq.~(\ref{eq:HTL-a}), the only effect of the chemical potential, once all excitation energies are measured with respect to the chemical potential,  being to change the plasmino frequency into $\omega_0=(g/4\pi)\sqrt{\mu^2+\pi^2T^2}$.

We now  proceed further, in the regime of ultra-soft momenta, $p\ll {\text{max}}(g^2T,g^2\mu)$. 
Since  the dominant contribution comes from the region $k\sim {\rm max}(T, \mu)$, the term $2k\cdot\tilde p$ in the denominator of Eq.~(\ref{eq:sigma-mu}) is small compared to $\deltam$ and one can expand in powers of $k\cdot\tilde p/\deltam$. The first two terms yield
\beq
\frac{1}{\deltam+2k\cdot \tilde{p}}\approx \frac{1}{\deltam}\left (1-\frac{2k\cdot \tilde{p}}{\deltam}\right).
\eeq
We write the corresponding contributions to the retarded self-energy as $\varSigma^\R=\varSigma_{(0)}^\R+\varSigma_{(1)}^\R$. We have
\beq\label{Sigma0R}
\varSigma_{(0)}^\R=\gamma^0\const \mu,
\eeq
where
\begin{align}
\label{eq:A}
\const&\equiv  \frac{g^2}{4 \pi^2 \mu\deltam} I_0(T,\mu),\\
\nonumber
I_0(T,\mu)&\equiv  \int_0^\infty {\rm d} |\vk| \, |\vk|^2\sum_{s=\pm 1}\left[ n_F(s|\vk|)+n_B(s|\vk|)  \right]\\
\label{eq:I0}
&= \frac{\mu}{3}(\pi^2T^2+\mu^2).
\end{align}
Here we have neglected vacuum terms, i.e., terms that do not depend on neither $T$ nor $\mu$.
The first order contribution can be put in the form
\beq\label{Sigma1R}
\varSigma_{(1)}^\R(p)=  - \frac{(p^0+i\damping)\gamma^0+v\,\vp\cdot\vgamma }{\Z}, 
\eeq
with $v= 1/3$, and 
\beq
\label{eq:Z}
\frac{1}{\Z}=\frac{g^2}{2\pi^2(\deltam)^2} \,I_1(T,\mu),
\quad \damping= \frac{I_2(T,\mu)}{I_1(T,\mu)}.
\eeq
Note that the value $1/3$ of the velocity is identical to that for the plasmino. It results from angular integration, and it is independent of the coupling strength.  
The integrals $I_1$ and $I_2$ are given by
 \begin{align}
I_1(T,\mu)&\equiv  \int^\infty_0 d|\vk| |\vk|^3 \sum_{s=\pm 1}s(\nf(s|\vk|)+\nb(s|\vk|)) \nonumber\\
\label{eq:I1}
&= \frac{1}{4}(\pi^2T^2+\mu^2)^2,\\
\label{eq:I2}
I_2(T,\mu)&\equiv  \int^\infty_0 d|\vk| |\vk|^3 \sum_{s=\pm 1}s(\nf(s|\vk|)+\nb(s|\vk|))\zeta(|\vk|).
\end{align}
Combining the two contributions of Eqs.~(\ref{Sigma0R}) and (\ref{Sigma1R}), one can write the fermion retarded self-energy as follows
\begin{align}
\begin{split} 
\varSigma^\R(p)&\approx \gamma^0\const \mu
- \frac{(p^0+i\damping)\gamma^0+v\vp\cdot\vgamma }{\Z},
\end{split}
\end{align}
from which one gets
\beq
\label{eq:expansion-ab}
a(p^0,|\vp|)=A\mu-\frac{p^0+i\varGamma}{Z},
\qquad  b(p^0,|\vp|)=-\frac{v |\vp|}{Z}.\nonumber\\
\eeq
The corresponding fermion retarded propagator reads
\begin{align}\label{retardedpropa}
\begin{split} 
S^\R(p)&= -\frac{1}{\Slash{p}-\varSigma^\R(p)} \\
&\approx \frac{1}{\varSigma^\R(p)}\\
&\approx -\frac{\Z}{2}
\Bigl(\frac{\gamma^0-\vgamma\cdot\hat{\vp}}{p^0+v|\vp|-\Z \const\mu+i\damping} \\
&~~~+ \frac{\gamma^0+\vgamma\cdot\hat{\vp}}{p^0-v|\vp|-\Z \const\mu+i\damping} \Bigr),  
\end{split}
\end{align}
where, in the second line, we have used the fact\footnote{This inequality is of course not valid in the vicinity of the quasi-goldstino pole. However, it is easily verified that keeping the free term in the pole condition only modifies Eq.~(\ref{pole condition}) by a negligible amount.} that $p\ll \varSigma^\R(p)\sim \mu, p/\cp^2$. 
This expression of the retarded propagator seems to suggest the existence of a pole at the  position
\begin{align}\label{pole condition}
\begin{split}
p^0&= \mp v|\vp|+\Z \const\mu-i\damping.
\end{split}
\end{align}
The quantity $Z$ can be interpreted as the residue at the pole by using Eq.~(\ref{retardedpropa}). 
However we need to make sure that this pole is located in the region where the calculation leading to Eq.~(\ref{retardedpropa}) is justified. This is the analysis to which we proceed now.

\subsection{$T\neq 0$ and $\mu= 0$ case}

To get a clear contrast to the case of finite $\mu$, we write first the result in the case of $T\neq 0$ and $\mu= 0$, which was investigated in Ref.~\cite{QED-ultrasoft}.
This is the case where charge conjugation symmetry holds.  Then it is easily verified that $I_0=0$. To see that, note that
\beq
n_B(-|\vk|)+n_B(|\vk|)=-1,\quad  n_F(|\vk|)+n_F(-|\vk|)=1,\nonumber\\
\label{eq:cancellation}
\eeq
where the second equality holds only if $\mu=0$. 
The resulting cancellation of the statistical factors is identical to the cancellation already discussed after Eq.~(\ref{eq:selfenergy-expression}). 
The pole occurs at $p^0= \mp v|\vp|-i\damping$, and corresponds to a very weakly damped mode. 
Since $\zetaf$, $\zetab\sim \cp^4T$ in the present case, we get $\damping\sim \cp^4T$ by using Eqs.~(\ref{eq:Z}) and (\ref{eq:I2}).
The residue is given by 
\beq
Z=\frac{g^2}{72\pi^2}.
\eeq
As already mentioned the ultra-soft fermionic excitation carries a very small spectral weight. 

\subsection{$T=0$ and $\mu\neq 0$ case}

Consider now the other extreme case, where $T=0$ and $\mu\neq 0$. 
In this case, setting $T=0$ in Eqs.~(\ref{asymmasses}), ~(\ref{eq:I0}), and (\ref{eq:I1}), we get
\beq
  A=\frac{2}{9},\quad Z=\frac{9g^2}{8\pi^2}
\eeq
from Eqs.~(\ref{eq:A}) and (\ref{eq:Z}).
The non vanishing of $A$ is intimately related to the breaking of charge conjugation symmetry by the chemical potential term. 
Because of this term, $\varSigma^R(p^0, \vzero)$ is no longer an odd function of $p^0$, as can be seen from Eq.~(\ref{eq:expansion-ab}). 
As we have also argued earlier, the breaking of charge invariance is accompanied by an explicit breaking of the supersymmetry. 
It is then an interesting question to see whether the ultra-soft fermionic excitation continues to exist in these conditions. 
We shall see that this is not the case. 

\begin{figure}[t] 
\begin{center}
\includegraphics[width=0.5\textwidth]{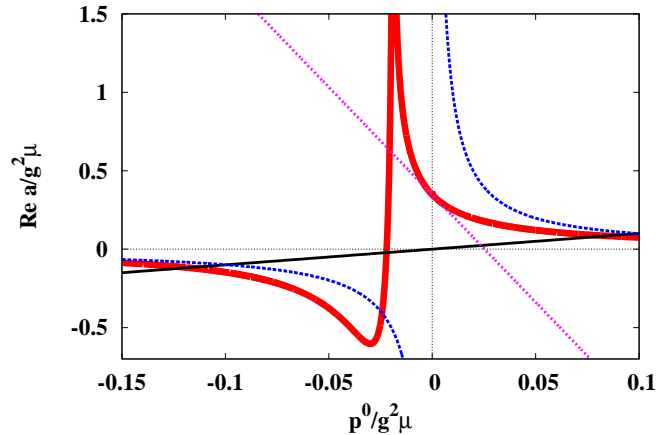} 
\caption{The real part of $a(p^0,0)$ as a function of energy ($p^0$) at  $T=0$, $\mu\neq 0$, and $\cp=0.8$ in the HDL approximation~\cite{HDL} (dashed, blue), in the resummed one-loop approximation with $\zeta=0$ (thick solid, red), Eq.~(\ref{eq:sigma-mu}), and in the small $p^0$ expansion with $\damping=0$ (dotted, magenta), Eq.~(\ref{eq:expansion-ab}).
The pole position is determined by the intersection of the curves and the inverse free propagator (solid, black) $p^0$.
The pole corresponding to the plasmino exists.
By contrast, there is no pole corresponding to the ultra-soft excitation, due to the fact that the resummed perturbation result departs very quickly from the small $p^0$ expansion result. 
There is also another solution, but it is not physical  since the corresponding crossing point occurs in a region where the slope of $a$  is positive.
}  
\label{fig:selfenergy-mu1}
\end{center}
\end{figure}

Indeed, because $\Z \const\mu\sim\cp^2\mu$, the pole would occur in a region that is not compatible with  the assumption $p\ll \cp^2\mu$, at the basis of our expansion. In fact, as shown in Fig.~\ref{fig:selfenergy-mu1} the actual self-energy departs very quickly from its linear approximation, and the only surviving pole is that of the plasmino.
 
The present situation is that of maximal violation of charge conjugation symmetry\footnote{It also corresponds to a large explicit breaking of supersymmetry. We distinguish explicit supersymmetry breaking associated with the term proportional to $\mu$ in the Lagrangian (\ref{eq:lagrangian}), from the spontaneous breaking associated with different finite masses for fermions and bosons coming from interaction effects.}. 
Actually, in terms of $\varLambda_{\pm}$, the present analysis is described as
\begin{align}
\label{eq:kinetic-mu}
\left(2iv\cdot\partial_X+\frac{\delta m^2}{|\vk|}\right)\varLambda_+(\vk,X)
&=\cp\Slash{v}\theta(\mu-|\vk|)\varPsi(X)
\end{align}
and $\varLambda_-=0$, while $|\varLambda_+|=|\varLambda_-|$ in the charge conjugation symmetric case with $\mu=0$.
As was explained in Sec.~\ref{sec:ultrasoft}, it is essential for the existence of the goldstino that the (hard) fermion and anti-fermion contributions to the process in Fig.~\ref{fig:kinetic} cancel, i.e., that $\varLambda_++\varLambda_-=0$, leading to the  pole condition $a(p^0, 0)=0$ (at $|\vp|=0$).
This cancellation is guaranteed by the charge conjugation symmetry.

It is interesting to note here that, in contrast to what happens to the goldstino,  the existence of the plasmino is not affected  by this breaking of charge symmetry. This is because, as explained in detail in Ref.~\cite{blaizot-mu}, an apparent particle-antiparticle symmetry develops at large chemical potential. This arises from the fact that the processes responsible for the existence of the plasmino involve (hard) fermions at the top of the Fermi sea, and holes at the bottom of the Fermi sea. But, if the fermion mass is small compared to the chemical potential (more precisely compared to $g\mu$), there is no difference between a hole at the bottom of the  Fermi sea, and a hole at the top of the Dirac sea, hence the apparent particle-antiparticle symmetry for the soft excitations. It is this symmetry that guarantees that $a(p_0,0)$ remains an odd function of $p_0$, in spite of the breaking of charge conjugation symmetry (which was used to establish this result for $\mu=0$ in the beginning of Sec.~\ref{sec:ultrasoft}). 
With energies measured with respect to the chemical potential,  the plasmino pole condition reads $p^0-a(p^0, 0)=0$, with $a(-p^0, 0)=-a(p^0, 0)$, and both $p^0$ and $a$ of order $\cp \mu$.

We shall  next consider the intermediate situation where the explicit symmetry breaking is sufficiently small in order not to alter the pattern that emerges from the spontaneous breaking of supersymmetry, and in particular allow for the existence of the quasi-goldstino. 

\begin{figure}[t] 
\begin{center}
\includegraphics[width=0.5\textwidth]{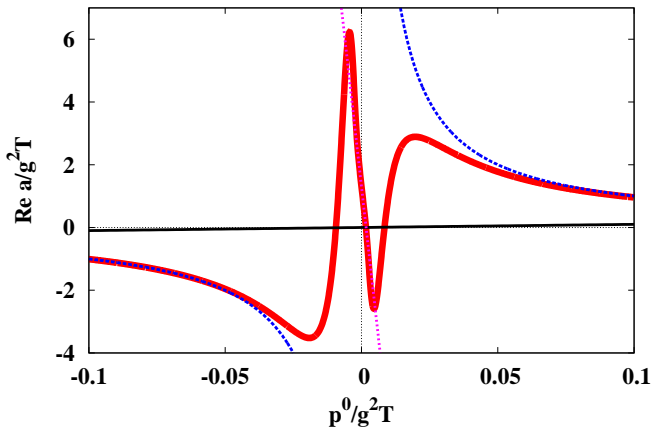} 
\caption{ 
The real part of $a(p^0,0)$ as a function of energy ($p^0$) at  $\mu=0.5T$, and $\cp=0.8$ in the HTL/HDL approximation (dashed, blue), in the resummed one-loop approximation with $\zeta=0$ (thick solid, red), from Eq.~(\ref{eq:sigma-mu}), and in the small $p^0$ expansion with $\damping=0$ (dotted, magenta), Eq.~(\ref{eq:expansion-ab}).
The pole position is determined by the intersection of the curves and the inverse free propagator (solid, black) $p^0$.
There is a pole corresponding to an ultra-soft excitation near $p^0=0$.  
} 
\label{fig:selfenergy-mu2}
\end{center}
\end{figure}

\begin{figure}[t] 
\begin{center}
\includegraphics[width=0.5\textwidth]{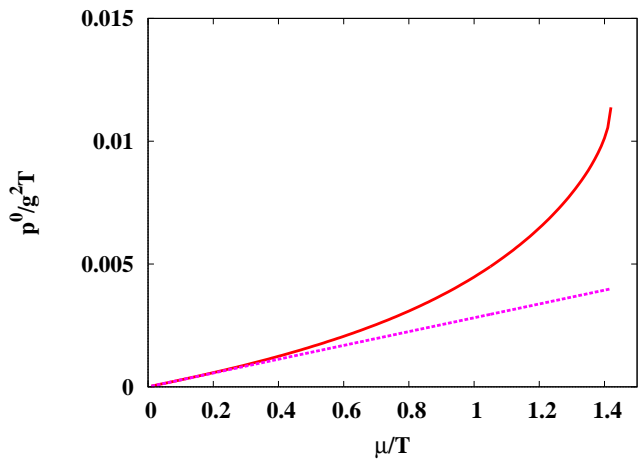} 
\caption{The pole $p^0$ as a function of the chemical potential $\mu$ at $\cp=0.8$, in the resummed one-loop approximation with $\zeta=0$ (solid, red), from Eq.~(\ref{eq:sigma-mu}), and in the small $p^0$ expansion with $\damping=0$ (dotted, magenta), Eq.~(\ref{eq:expansion-ab}).
}
\label{fig:pole-mu}
\end{center}
\end{figure}

\subsection{$T\gtrsim\mu\neq 0$ case}
\label{sec:T-mu}

Consider first the case $T\gg\mu$. In this case, we have
\beq
\deltam&\approx\frac{g^2T^2}{24}, \quad
I_0(\mu,T)\approx \frac{\pi^2T^2}{3}\mu,\quad
I_1(\mu,T)&\approx \frac{\pi^4T^4}{4}.\nonumber\\
 \eeq
One then gets
\beq
\const = 2,\quad \Z=\frac{g^2}{72\pi^2}.
\eeq
The pole condition is the same as before, Eq.~(\ref{pole condition}), namely
\begin{align}
\label{eq:pole-Tmu}
\begin{split} 
p^0&= \mp v|\vp|+\Z \const\mu-i\damping .
\end{split}
\end{align}
The damping rates of the hard particles are $\zetaf$, $\zetab\sim \cp^4T$, so that $\damping\sim\cp^4T$, as in the case of $T\neq 0$ and $\mu=0$. 
In contrast to the case of $T=0$ and $\mu\neq 0$, now we see that the pole position, Eq.~(\ref{eq:pole-Tmu}), does not break the condition of the expansion, $\tilde{p}\ll \cp^2T$, since $\Z A\mu\sim \cp^2\mu\ll\cp^2T$. 
This is confirmed in Fig.~\ref{fig:selfenergy-mu2}, corresponding to $T=2\mu$:
There is a crossing point near $p^0=0$,  and around that crossing point,  the full result of  resummed perturbation theory is almost the same as that of the small $p^0$ approximation. Thus, the quasi-goldstino continues to exist at  finite chemical potential when the condition $T\gg\mu$ is satisfied, that is, as long as charge conjugation\footnote{We also note that the existence of the quasi-goldstino is affected by the breaking of chiral symmetry caused by finite bare fermion mass~\cite{lebedev,mitsutani}.  
On the other hand, the quasi-goldstino survives when massless fermions interact with a massive boson, which does not break the chiral symmetry, provided again the boson mass is not too large~\cite{3peak}.} and/or supersymmetry are not too strongly violated. 

\begin{figure}[t] 
\begin{center}
\includegraphics[width=0.5\textwidth]{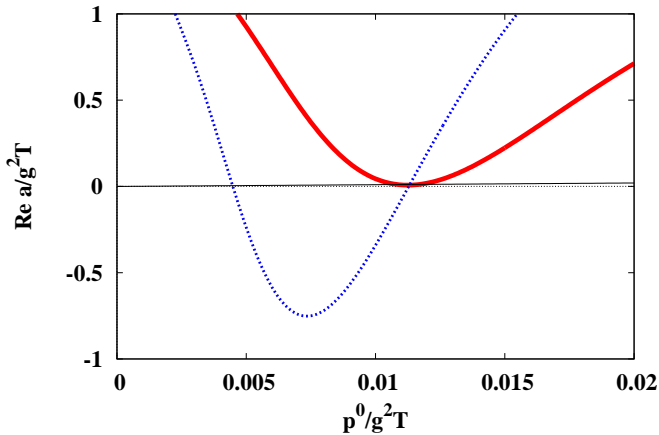} 
\caption{
The real part of $a(p^0,0)$ as a function of energy ($p^0$) in the resummed one-loop approximation Eq.~(\ref{eq:sigma-mu}) with $\zeta=0$ at $\cp=0.8$, for $\mu=1.415T$ (thick solid, red), which is near $\mu_{\text{max}}$.
The pole position is determined by the intersection of the (negative slope parts of the) curves and the horizontal, $p^0$ axis.
We see that the slope is approximately zero at the pole.
For contrast, we also plotted Re $a(p^0,0)$ for $\mu=T$ (dotted, blue), which has finite slope at the pole.
} 
\label{fig:slope}
\end{center}
\end{figure}

When $\mu$ is comparable to $T$, we need to perform a numerical evaluation of Eq.~(\ref{eq:sigma-mu})  to see the fate of the ultra-soft fermion mode. 
As a result of such an evaluation with $\zeta=0$ and $0<\cp<1.0$, we find that the maximum value of the chemical potential ($\mu_{\text{max}}$) for which the quasi-goldstino exists is  $\mu_{\text{max}}=1.41T$, regardless of the value of $\cp$.
The independence on $g$  is easy to understand:
By  introducing the dimensionless quantities, $k'\equiv |\vk|/T$, $\mu'\equiv\mu/T$, $p'{}^0\equiv p^0/(\cp^2T)$, and $\delta m'{}^2\equiv \delta m^2/(g^2T^2)$, one gets from Eq.~(\ref{eq:sigma-mu}) at $|\vp|=0$ (and $\zeta=0$)
\begin{align}
\label{eq:selfenergy-dimensionless}
\begin{split} 
a(p^0,0)&= \frac{T}{4\pi^2}\int^\infty_0 dk' \sum_{s=\pm 1} \\
&~~~\times\left(\frac{1}{e^{k'-s\mu'}+1}+\frac{1}{e^{k'}-1}\right)
\frac{sk'{}^2}{\delta m'{}^2+2sk' p'{}^0},
\end{split}
\end{align}
by neglecting $T=\mu=0$ part.
We note that $a/T$ is a function of $\mu'$ and $p'{}^0$ with no dependence on $g$. We set $a'(\mu', p'{}^0)\equiv a(p^0,0)/T$ from now on.
The pole condition (for $p^0\ll a$), $a(p_0,0)=0$,  becomes 
\begin{align}
\label{eq:pole-dimensionless}
a'(\mu', p'{}^0)&=0, 
\end{align}
 so whether the pole exists or not does not depend on $\cp$.
Therefore, $\mu_{\text{max}}$ does not, either depend on $g$.
Equation (\ref{eq:selfenergy-dimensionless}) also indicates that the pole position is proportional to $\cp^2$, since  $p^0= \cp^2T p'{}^0$ is proportional to $\cp^2$. 
The numerical evaluation of Eq.~(\ref{eq:sigma-mu}) with $\zeta=0$ confirms this dependence of the pole on $\cp^2$  for any $\mu<\mu_{\text{max}}$.
Finally, we plot the pole as a function of $\mu$ at $\cp=0.8$ in Fig.~\ref{fig:pole-mu}.
We see that, when $\mu$ is small compared with $T$, the pole calculated from Eq.~(\ref{eq:sigma-mu}) agrees well with that calculated from Eq.~(\ref{eq:expansion-ab}).
This behavior is consistent with our expectation obtained in the first part of this subsection analytically.
When $\mu$ is large, the difference of the two poles become significant. In particular,   the slope of the curve giving the energy of the mode as a function of $\mu$ seems to diverge at $\mu=\mu_{\text{max}}$.
This behavior can be related to that  of the function $a(p^0)$ displayed in Fig.~\ref{fig:slope}. Indeed, by
 taking the derivative of Eq.~(\ref{eq:pole-dimensionless}) with respect to $\mu'$, we get
\begin{align}
\frac{\partial a'}{\partial \mu'}+\frac{\partial p'{}^0}{\partial \mu'}\frac{\partial a'}{\partial p'{}^0}&=0, 
\end{align}
at the pole, $p'{}^0(\mu')$.
Around $\mu=\mu_{\text{max}}$, $\partial a'/(\partial p'{}^0)$ approaches zero, as can be seen from Fig.~\ref{fig:slope}, so as long as $\partial a'/(\partial \mu')$ is finite, $\partial p'{}^0/(\partial \mu')$ should diverge.


\section{Summary and Concluding Remarks}
\label{sec:summary}

We analyzed the fermion retarded propagator in a system of massless fermions and scalar bosons interacting through a simple Yukawa coupling with strength $\cp$. 
We focused on the particular regime of ultra-soft momenta $\cp^2T$ or $\cp^2\mu$, and studied the conditions under which a collective excitation can develop.  
Following Ref.~\cite{lebedev}, we  linked the origin of the ultra-soft excitation to the spontaneous breaking of a supersymmetry of the non interacting system, caused by the difference between the thermal masses of the high momentum fermion and boson; hence the name quasi-goldstino given to the ultra-soft excitation.
 We also emphasized the role of charge conjugation symmetry. The chemical potential breaks both charge conjugation and supersymmetry (explicitly), and this breaking is sufficiently strong to prevent the existence of the quasi-goldstino at zero temperature. 
However, we verified   that the quasi-goldstino exists in the high temperature, finite fermion density, system as long as $T \geq 0.71\mu$.

The analysis in this paper could be generalized  to the more interesting cases of QED/QCD at finite density, with the extra complication that the damping rates are anomalously large then, and vertex corrections need to be taken into account. 
It is unlikely though that such fine details of the fermion spectrum could be observed in relativistic plasmas, such as for example the quark-gluon plasmas produced in heavy ion collisions. However, we note that cold atoms offer interesting possibilities to prepare systems~\cite{Yu:2007xb} in which some of the phenomena discussed in the present paper can be realized. 
This will be the object of a forthcoming publication~\cite{future}. 

\section*{Acknowledgement}

D. S. thanks Y. Hidaka for fruitful discussions and comments.
D. S. was supported by JSPS KAKENHI Grant Numbers 24$\cdot$56384, the Grant-in-Aid for the Global COE Program ``The Next Generation of Physics, Spun from Universality and Emergence'' from the Ministry of Education, Culture, Sports, Science and Technology (MEXT) of Japan, and JSPS Strategic Young Researcher Overseas Visits Program for Accelerating Brain Circulation (No. R2411). 
JPB's research is supported by the European Research Council under the Advanced Investigator Grant ERC-AD-267258.


\end{document}